\documentclass[prb,twocolumn,groupedaddress,showpacs]{revtex4}

\usepackage{graphicx}
\usepackage{dcolumn}
\usepackage{amssymb}
\usepackage{bm}
\usepackage{epsfig}

\begin{document}

\title{Parity meter for charge qubits: an efficient quantum entangler}

\author{B.~Trauzettel$^{1,2}$, A.~N.~Jordan$^{3,4}$, C.~W.~J.~Beenakker$^1$, and M.~B{\"u}ttiker$^3$}
\affiliation{${}^1$Instituut-Lorentz, Universiteit Leiden, P.O.
Box 9506,
2300 RA Leiden, The Netherlands\\
${}^2$Kavli Institute of Nanoscience, TU Delft, P.O. Box 5046, 2600 GA Delft, The Netherlands\\
${}^3$D{\'e}partement de Physique Th{\'e}orique, Universit{\'e} de
Gen\`eve, 1211 Gen\`eve 4, Switzerland \\
${}^4$Institute for Quantum Studies, Texas A\&M University,
College Station, Texas 77843-4242, U.S.A.}

\date{April 2006}

\begin{abstract}
We propose a realization of a charge parity meter based on two
double quantum dots alongside a quantum point contact. Such a
device is a specific example of the general class of mesoscopic
quadratic quantum measurement detectors previously investigated by
Mao {\it et al.} [Phys. Rev. Lett. {\bf 93}, 056803 (2004)]. Our
setup accomplishes entangled state preparation by a current
measurement alone, and allows the qubits to be effectively
decoupled by pinching off the parity meter. Two applications of
the parity meter are discussed: the measurement of Bell's
inequality in charge qubits and the realization of a controlled
NOT gate.
\end{abstract}


\maketitle

\section{Introduction}

Recently, it has been realized that parity meters based on solid
state structures should be very promising candidates to create
entanglement of electronic systems. \cite{beena04,engel05,egues05}
Whereas previous proposals and applications of parity meters in
the solid state have dealt with the entanglement of the spin
degree of freedom, we presently investigate a parity meter based
on charge qubits, which is able to generate entanglement in a
solid state system just by measuring a DC current through a
quantum point contact (QPC). The setup under consideration,
schematically shown in Fig.~\ref{pmeter} {\bf (A)}, is a
particular example of the general class of mesoscopic quadratic
quantum measurement devices investigated by Mao {\it et al.} in
Ref.~\onlinecite{mao04}. The design of our device has been
inspired by the work of Ruskov and Korotkov, where it has been
demonstrated that current and noise measurements of a QPC coupled
to two charge qubits can be used as an entangler. \cite{rusko03}
We propose a setup, in which this task can be achieved by a
current measurement only.

The charge qubits, e.g. a single electron in a tunnel-coupled
double quantum dot (DQD), are coupled capacitively to the
measurement device (the parity meter). Recently, coherent quantum
oscillations have been measured in DQD systems. \cite{hayas03}
Quantum detectors based on QPC's coupled to DQD qubit systems have
been studied intensively in the past, both theoretically
\cite{dqd_theo} as well as experimentally. \cite{dqd_exp}
Transport properties of Coulomb coupled DQD systems have also been
analyzed. \cite{lambe06}

Generally speaking, the idea of a parity meter is that it can
distinguish between the subspaces of two parity classes of quantum
states but it cannot distinguish between different states in each
parity class. It has been demonstrated that such a device can be
used to implement a controlled NOT (CNOT) gate. \cite{beena04} A
CNOT gate is a universal quantum gate, and therefore enables
universal quantum computation when combined with single qubit
gates.

The design of the parity meter that we propose in this article is
very straightforward. It just relies on two qubit systems (based
on DQD's) and a single measurement device (based on a QPC). All
elements of the parity meter can be built with standard
lithographic techniques in the lab. If we think about these qubits
as DQD's in GaAs heterostructures then dephasing due to coupling
of the charge on the dots to acoustic phonons \cite{voroj05} and
dephasing due to background charge fluctuations \cite{itaku03}
cause severe problems. Nevertheless, charge qubits based on DQD's
in other structures such as carbon nanotubes \cite{mason04} or
semiconductor nanowires might have much better coherence
properties (due to the fact that they are essentially one
dimensional systems), which would make our predictions measurable.
Recently, charge qubit operations of an isolated (leadless)
silicon double quantum dot have been reported with an extremely
long coherence time. \cite{gorma05}

The article is organized as follows. In Sec.~\ref{sec_trans}, we
propose a specific realization of a charge parity meter, i.e. a
quadratic quantum measurement device. Subsequently, in
Sec.~\ref{pent_sec}, we demonstrate how the parity meter acts as
an entangler. In Sec.~\ref{sec_bell}, we discuss two applications
of the parity meter: (i) the measurement of a violation of Bell's
inequality and (ii) a realization of a CNOT gate. We conclude in
Sec.~\ref{sec_con}.

\section{Realization of a charge parity meter}
\label{sec_trans}

\begin{figure}
\vspace{0.3cm}
\begin{center}
\epsfig{file=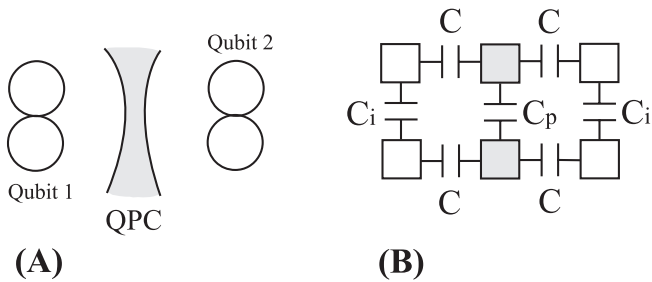,scale=0.55,=}
\caption{\label{pmeter} {\it Parity meter setup.} In {\bf (A)},
two DQD's are shown with a QPC in between them. By measuring the
current through the QPC, we are able to determine the parity class
of the two-qubit state formed by a single electron in each DQD
qubit. In {\bf (B)}, a capacitive model is illustrated for the
setup considered in {\bf (A)}. The two grey boxes in the middle
correspond to the dipole across the QPC.}
\end{center}
\end{figure}

It has been pointed out in Ref.~\onlinecite{mao04} that a general
quadratic quantum measurement device provides a simple way of
entangling two otherwise noninteracting qubits. In such a
measurement device, the transmission amplitude $t$ of some
particles, e.g. electrons, should depend on the measurement basis
operators $\sigma_z^{(\alpha)}$ of the two qubits $\alpha = 1,2$ in
the following way
\begin{equation} \label{quadc}
t \propto \sigma_z^{(1)} \sigma_z^{(2)} .
\end{equation}
In this section, we demonstrate that a physical realization of a
parity meter for charge qubits consists of two DQD's and a single
QPC. The setup we have in mind consists of two DQD's alongside a
QPC, {\it cf.} Fig.~\ref{pmeter} {\bf (A)}. We will show that a
capacitance model based only on {\it linear} relations between
charge and potentials explains the existence of the desired {\it
quadratic} coupling (\ref{quadc}) in our device. While in
Ref.~\onlinecite{mao04} the quadratic measurement is achieved by
considering an inflection point of the transmission with
$dt/dU=0$, where $t$ is the transmission amplitude and $U$ the
potential of the measurement device, the quadratic coupling which
we discuss below is realized for arbitrary transmission since it
is a consequence of the spatial symmetry of our arrangement. A
similar idea, using geometric symmetry (in a more complicated
interferometric structure) to realize a quadratic detector has
been proposed in Ref.~\onlinecite{rusko06}.

We assume that each DQD contains a single electron, which
therefore acts as a charge qubit, and can be described by the
Hamiltonian
\begin{equation}
H_{\rm QD, \alpha} = -\frac{1}{2} \Bigl( \epsilon_\alpha
\sigma_z^{(\alpha)} + \Delta_\alpha \sigma_x^{(\alpha)} \Bigr) .
\end{equation}
Here $\alpha = 1,2$ (for the two different qubits),
$\epsilon_\alpha$ is the difference of single particle energy
levels in each dot, $\Delta_\alpha$ is the tunnel coupling between
the dots, and $\sigma^{(\alpha)}_{x}$ is a Pauli matrix acting on
qubit $\alpha$. The direct electrostatic coupling between the two
DQD's is neglected. However, the two qubits are indirectly coupled
to each other via the QPC.

The corresponding coupling term, in our setup, may be written as
\begin{equation} \label{hint}
H_{\rm int} = (\Delta \hat{E}/2) \sigma_z^{(1)} \sigma_z^{(2)} ,
\end{equation}
where $\Delta \hat{E}$ is a charging energy operator with a
quantum expectation value $\Delta E =E_E - E_O \equiv \langle
\Delta \hat{E} \rangle$ equal to the difference in charging energy
of the even (index $E$) and odd (index $O$) parity class. In the
tunnelling regime, the operator $\Delta \hat{E}$ can be associated
with the standard detector input variable $\lambda a_R^\dagger a_L
+ \ {\rm H.c.} \ $, where $\lambda$ is the coupling constant,
$a_R^\dagger$ is the creation operator for an electron in the one
lead of the detector, and $a_L$ an annihilation operator in the
other lead. \footnote{We have derived the Hamiltonian (\ref{hint})
in the scattering picture of transport through the detector. On
the level of observables such as the average current and the
noise, the scattering picture of transport and the tunnelling
picture of transport are equivalent in the tunnelling regime. The
validity range of the scattering picture, however, also extend to
the regime of an open QPC.} The combined Hamiltonian of the two
DQD's, the QPC, and the coupling (\ref{hint}) then reads
\begin{equation}
H = H_{\rm QD,1} + H_{\rm QD,2} + H_{\rm QPC} + H_{\rm int} ,
\end{equation}
where $H_{\rm QPC}$ is the Hamiltonian of the detector.

To justify the interaction Hamiltonian (\ref{hint}) in the case of
an arbitrary two qubit parity detector, the coupling Hamiltonian
may be expressed in the basis, defined with each of the
configurations $\{\vert\uparrow \uparrow \rangle, \vert\downarrow
\downarrow \rangle,\vert\uparrow \downarrow \rangle,
\vert\downarrow \uparrow \rangle \}$, where we assign pseudo-spins
to the position of the electron in the DQD: spin $\uparrow$
corresponds to the case where the electron is in the upper dot and
spin $\downarrow$ corresponds to the case where the electron is in
the lower dot. The even parity class contains the states $\{
|\uparrow \uparrow \rangle, |\downarrow \downarrow \rangle \}$,
while the odd parity class contains $\{ |\uparrow \downarrow
\rangle, |\downarrow \uparrow \rangle \}$. We will show below that
the parity detector can distinguish only between the even and the
odd parity classes. This defines the basis that is accessible to
the parity detector.  The parity detector cannot distinguish the
states in either even or odd subclass, but it can distinguish
between parity subclasses. In particular, the energy of both even
configurations are equal, and the energies of both odd
configurations are equal, but the even energy is not equal to the
odd energy. Therefore the coupling Hamiltonian can be expressed as
\begin{eqnarray}
H_{\rm int} &=& \hat{E}_E (\vert\uparrow \uparrow \rangle \langle
\uparrow \uparrow \vert + \vert\downarrow \downarrow \rangle
\langle
\downarrow \downarrow \vert) \nonumber \\
&+& \hat{E}_O (\vert\uparrow \downarrow \rangle \langle\uparrow
\downarrow \vert + \vert\downarrow \uparrow \rangle \langle
\downarrow \uparrow \vert) .
\end{eqnarray}
Introducing the projection operators, ${\rm
P}^{(\alpha)}_{\uparrow, \downarrow}$ on the up or down state of
qubit $\alpha$, going to sum and difference variables,
$\hat{\Sigma} = (\hat{E}_E + \hat{E}_O)/2, \Delta \hat{E} =
(\hat{E}_E - \hat{E}_O)$, and recalling that ${\rm
P}^{(\alpha)}_{\uparrow} + {\rm P}^{(\alpha)}_{\downarrow}={\bf
1}^{(\alpha)},\, {\rm P}^{(\alpha)}_{\uparrow} - {\rm
P}^{(\alpha)}_{\downarrow}=\sigma_z^{(\alpha)}$, we find that the
coupling Hamiltonian may be rewritten as
\begin{equation}
H_{\rm int} = \hat{\Sigma} \, {\bf 1}^{(1)} {\bf 1}^{(2)} +
(\Delta \hat{E}/2) \sigma_z^{(1)} \sigma_z^{(2)}.
\label{genparity}
\end{equation}
The first term may be absorbed into the detector Hamiltonian, and
the second term recovers our coupling Hamiltonian (\ref{hint}).

We will now demonstrate that a single QPC placed in a special way
between two DQD's has a Hamiltonian of the form (\ref{hint}). As
illustrated in Fig.~\ref{pmeter} {\bf (B)}, we divide the two
DQD's as well as the QPC in two regions with two corresponding
charges each. The left DQD has charge $Q_1 = - Q_2 = Q_L$ in dots
with potentials $U_1$ and $U_2$. The right DQD has charge $Q_3 = -
Q_4 = Q_R$ in dots with potentials $U_3$ and $U_4$. We set $e
\equiv 1$ and $\hbar \equiv 1$. Suppose that, in the odd parity
class with $Q_L Q_R = -1$, the QPC has a potential $V(x)$ along
the $x$-axis given by \cite{butti90}
\begin{equation}
V(x) = V_0 - \frac{1}{2} m \omega_x^2 x^2 + {\cal O} (x^4) .
\end{equation}
We are interested in the change of the potential $V_0$ at the
saddle point as the qubit configuration changes from the odd to
the even parity class. We call the saddle point potential of the
even parity class $V_1$. The saddle point potentials allow us to
determine the transmission coefficient of the QPC.

As compared to the odd parity class, the even class will polarize
the QPC, i.e. there will be an additional electrical field $E$ at
the saddle point with a potential $V(x) = V_0 + e E x -
\frac{1}{2} m \omega_x^2 x^2 + \dots$, which will give rise to a
shift in the location of the saddle point and an increase in its
height
\begin{equation} \label{v1}
V_1 = V_0 + \frac{e^2 E^2}{2m\omega_x^2} .
\end{equation}
To estimate the field $E$ we consider the capacitive model shown
in Fig.~\ref{pmeter} {\bf (B)}. The QPC dipole is described with a
capacitance $C_p$ and the dipole charges $Q_5 = Q_d$ and $Q_6 =
-Q_d$ exist in regions with potentials $U_5$ and $U_6$.
\cite{chris96} A charge on a dot next to a dipole region has a
capacitance $C$ and the interaction of the charges across a DQD is
described by an (internal) capacitance $C_i$. For simplicity, we
assume that only nearest neighbor charge configurations couple to
each other. This is justified because the QPC screens the DQD's,
and, therefore, decreases the direct coupling between the DQD's
substantially. We now have for the DQD charges the equations
\begin{eqnarray}
Q_1 &=& C_i(U_1-U_2) + C(U_1-U_5) = Q_L , \nonumber \\
Q_2 &=& C_i(U_2-U_1) + C(U_2-U_6) = - Q_L, \nonumber \\
Q_3 &=& C_i(U_3-U_4) + C(U_3-U_5) = Q_R , \nonumber \\
Q_4 &=& C_i(U_4-U_3) + C(U_4-U_6) = - Q_R,
\end{eqnarray}
and for the QPC dipole charges
\begin{eqnarray}
Q_5 &=& C_p(U_5-U_6) + C(U_5-U_1) + C(U_5-U_3) = Q_d, \nonumber \\
Q_6 &=& C_p(U_5-U_6) + C(U_5-U_1) + C(U_5-U_3) = - Q_d. \nonumber
\\
\end{eqnarray}
These become a complete set of equations if we assume that the
region over which the dipole extends has a density of states $D$
such that a small variation of the potential in these regions
gives rise to a charge $Q_5 = - D U$ and $Q_6 = D U$ with $U=(U_5
- U_6)/2$. Here we have assumed that the QPC potential is
spatially symmetric in the odd parity configuration. This requires
that the QPC is located symmetrically in between the DQD's and
permits us to take the density of states to the right and the left
of the QPC to be equal to $D \equiv D_L = D_R$.

We obtain for the Coulomb energy
\begin{eqnarray}
E_c &=& \frac{1}{2} \sum_i Q_i U_i \\
&=& \frac{Q_L(Q_L+CU)+Q_R(Q_R+CU)}{(2C_i+C)} - D U^2, \nonumber
\end{eqnarray}
where
\begin{eqnarray}
U &\equiv& U_5 = \frac{C}{C^2_{\rm sum}} (Q_L + Q_R)  \label{u} ,
\\
C^2_{\rm sum} &=& (2C_p + 2C +e^2 D)(2C_i + C) -2 C^2 \label{csum}
.
\end{eqnarray}
We see that there exist a contribution to the Coulomb energy
proportional to $Q_L Q_R$ given by
\begin{equation} \label{ec}
\Delta E_c = \frac{C^2}{C^4_{\rm sum}} \frac{(2C_i +C)C_p + 2 C_i
C}{(2C_i+C)} 4 Q_L Q_R .
\end{equation}
This contribution affects the saddle point potential in such a way
that the QPC acts as a parity meter. Eq.~(\ref{ec}) can be
identified with Eq.~(\ref{hint}).

\begin{figure}
\vspace{0.3cm}
\begin{center}
\epsfig{file=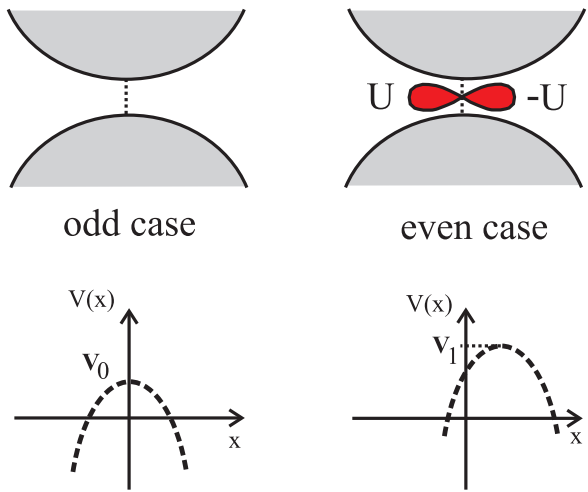,scale=0.85,=} \caption{\label{dipol} {{\it
Upper part:} Top view of QPC in the parity meter setup. In the odd
parity class, where $(Q_L,Q_R)$ is $(1,-1)$ or $(-1,1)$, no dipole
is generated across the QPC and it is nicely symmetric. In the
even parity class, where $(Q_L,Q_R)$ is $(1,1)$ or $(-1,-1)$, the
situation is different. There, a dipole is generated across the
QPC by the particular position of the electrons in the quantum
dots. The dipole shifts the saddle point to the right (left)
depending on the direction of the polarization of the dipole, but
for both polarizations the shift is to higher energies. Note that
the QPC in this figure is rotated by $90^\circ$ with respect to
the QPC in Fig.~\ref{pmeter}. {\it Lower part:} Illustration of
the saddle point potential for the two configurations.}}
\end{center}
\end{figure}

We now estimate the saddle point potential, in the even
configuration, which determines the transmission. The voltage drop
of the dipole is $2U$ and, assuming that the center of its charges
is separated by a distance $2d$, we find an electric field
$E=U/d$. Hence, using Eqs.~(\ref{v1}) and (\ref{u}) we obtain
\begin{equation}
V_1 = V_0 + \frac{1}{2m\omega_x d^2} \left( \frac{C}{C^2_{\rm
sum}} \right)^2 (Q_L + Q_R)^2 .
\end{equation}
Thus, the QPC has a saddle point height $V_0$ for the odd
configuration [$Q_L Q_R=-1$] and a somewhat higher saddle point
potential $V_1$ for the even configuration [$Q_L Q_R =1$]. We
illustrate in Fig.~\ref{dipol} the symmetric potential landscape
of the QPC in the odd case and the generated dipole across the QPC
in the even case as well as the corresponding saddle point
potentials.
The transmission probability $T=\vert t \vert^2$ of the QPC is
directly related to the saddle point height via \cite{butti90}
\begin{equation} \label{talp}
T_\alpha = \frac{1}{1+e^{-2 \pi (E_F-V_\alpha)/\omega_x}} ,
\end{equation}
where $E_F$ is the Fermi energy in the QPC, $\alpha = 0$ for the
odd case, and $\alpha=1$ for the even case. This shows that the
transmission probability through the system considered here has
indeed the desired property stated in Eq.~(\ref{quadc}).


To summarize, there are three physically distinguishable
situations:  (i) There is no charge dipole (both odd
configurations), (ii) there is a QPC charge dipole pointing up
(the even configuration $\vert \uparrow \uparrow \rangle$), and
(iii) there is a QPC charge dipole pointing down (the even
configuration $\vert \downarrow \downarrow \rangle$). Although all
three situations are distinguishable in principle, the crucial
fact is that the potential height experienced by the transport
electrons is the same in both even configurations. Therefore, the
current differs only between the even and the odd configuration.
For later reference, we define here the two currents
\begin{eqnarray}
I_{\rm odd} &=& \frac{2 e^2}{h} V_{\rm bias} T_0 , \\
I_{\rm even} &=& \frac{2 e^2}{h} V_{\rm bias} T_1 ,
\end{eqnarray}
where $V_{\rm bias}$ is the bias voltage across the QPC and we
have re-introduced $e$ and $\hbar$ for clarity. During the
measurement, the QPC should be operated in the regime of linear
$V_{\rm bias}$ but still $V_{\rm bias} \gg \Delta$. The reason is
that a large bias would break the symmetry between the $\vert
\uparrow \uparrow \rangle$ and the $\vert \downarrow \downarrow
\rangle$ state, so the parity meter could then distinguish the two
states of the even class. In typical systems (for instance the
ones investigated in Ref.~\onlinecite{dqd_exp}), $\Delta \approx
10 \; {\rm \mu eV}$, $V_{\rm bias} \approx 1 \; {\rm mV}$, and the
$I$-$V$ characteristics of the QPC is linear. Thus, the above
stated requirements can be easily met. The parity measurement time
\cite{dqd_theo}
\begin{equation}
T_M = \frac{4 S_I}{(I_{\rm odd} - I_{\rm even})^2} ,
\end{equation}
with $S_I = \int dt \langle \Delta I(t) \Delta I(0) \rangle$, and
$\Delta I(t) = I(t) - \langle I \rangle$ is the time scale
required to a obtain a signal-to-noise ratio of order 1. In
current (present day) GaAs-based quantum dot devices this time
scale
 will be of the order of a few $\mu s$ (see e.g.
Ref.~\onlinecite{vande04}) and therefore much longer than typical
coherence times of the order of a few $n s$. For quantum dots in
other physical systems such as carbon nanotubes or nanowires the
coherence times should be much longer and therefore the
measurement time $T_M$ can, in principle, become of the same order
as the coherence time or even smaller.

For the Bell measurement (discussed below), it is important that
the coupling between the two qubits (the two DQD's) can be turned
off after the state has been prepared. In our setup, the coupling
between the two DQD's is minimized by removing the bias voltage
($V_{\rm bias} = 0$) and by pinching off the QPC, therefore,
reducing $\Delta E$. For all operations that are done on a time
scale $\Delta t \ll 2/\Delta E$, the density matrix of the two
DQD's experiences little coherent evolution due to the coupling
term (\ref{hint}). Therefore, on such short time scales the two
qubits would behave as if they were decoupled.

\section{Parity meter as entangler}
\label{pent_sec}

At the symmetry point of the two-qubit system, where $\epsilon_1 =
\epsilon_2 = 0$ and $\Delta_1 = \Delta_2 \equiv \Delta$, our
device acts as an entangler just by measuring the current through
the QPC.

The Hamiltonian of the two qubits at the symmetry point (index SP)
reads
\begin{equation}
H_{\rm SP} = - \frac{\Delta}{2} \Bigl( \sigma_x^{(1)} +
\sigma_x^{(2)} \Bigr) .
\end{equation}
The eigenstates of $H_{\rm SP}$ are the two anti-symmetric Bell
states
\begin{eqnarray}
|\Psi_{B1} \rangle &=& \frac{1}{\sqrt{2}} \left( |\uparrow
\downarrow
\rangle - |\downarrow \uparrow \rangle \right) , \label{psib1} \\
|\Psi_{B2} \rangle &=& \frac{1}{\sqrt{2}} \left( |\uparrow
\uparrow \rangle - |\downarrow \downarrow \rangle \right) .
\label{psib2}
\end{eqnarray}
The state $|\Psi_{B1} \rangle$ belongs to the odd parity class and
the state $|\Psi_{B2} \rangle$ to the even one. Thus, both
eigenstates of the Hamiltonian $H_{\rm SP}$ can be distinguished
from each other by a current measurement through the QPC in
Fig.~\ref{pmeter}. The other two symmetric Bell states
\begin{eqnarray}
|\Psi_{B3} \rangle &=& \frac{1}{\sqrt{2}} \left( |\uparrow
\downarrow
\rangle + |\downarrow \uparrow \rangle \right) , \label{psib3}\\
|\Psi_{B4} \rangle &=& \frac{1}{\sqrt{2}} \left( |\uparrow
\uparrow \rangle + |\downarrow \downarrow \rangle \right)
\label{psib4},
\end{eqnarray}
are transformed into each other, obeying the time-evolution
\begin{equation} \label{statedyn}
|\Psi(t) \rangle = \cos(\Delta t + \phi) |\Psi_{B3} \rangle - i
\sin(\Delta t + \phi) |\Psi_{B4} \rangle .
\end{equation}
Here, $\phi$ is an arbitrary phase.

Since the states (\ref{psib1}) -- (\ref{psib4}) form a complete
basis of the two-qubit system, we can conclude that a measurement
of the current through the QPC can take three possible outcomes:
(i) $I_{\rm odd}$, which means that the system is driven into the
steady state (\ref{psib1}) by the measurement, (ii) $I_{\rm
even}$, which means that the system is driven into the steady
state (\ref{psib2}) by the measurement, and (iii) $I_{\rm mix}
\equiv (I_{\rm odd}+I_{\rm even})/2$ (with $I_{\rm odd} < I_{\rm
mix} < I_{\rm even}$), which means that the system is driven into
the state (\ref{statedyn}), which exhibits a dynamical detector
signal that is both oscillatory and noisy.

This behavior can be demonstrated within a simple model of
continuous parity measurement state preparation with a series of
projective parity measurements. Using a master equation
description of the time evolution of the density matrix of the
two-qubit system, we have verified that the latter model is in
agreement with continuous weak measurement. In the master equation
description, we have treated the electrons in the QPC as bath
variables and integrated out the bath degrees of freedom in the
weak coupling and Markovian regime. The resulting master equation
is of the Lindblad form, where the decay rates are proportional to
the auto-correlation function of the input variable of the parity
detector. The algorithm of our model of continuous parity
measurement state preparation is as follows:
\begin{enumerate}
\item{Fix the desired initial state in the measurement basis. It
can be randomly chosen, or can be fixed as a state that is
experimentally simple to prepare.} \item{Apply a unitary
transformation to change to the Bell basis, where Hamiltonian
evolution is simple. More explicitly, if we represent an arbitrary
state as
\begin{equation}
|\Psi \rangle = a |\downarrow \downarrow \rangle + b |\downarrow
\uparrow \rangle + c |\uparrow \downarrow \rangle + d |\uparrow
\uparrow \rangle ,
\end{equation}
a simple basis transformation enables us to write the same state
as
\[
|\Psi \rangle = \alpha |\Psi_{B1} \rangle + \beta |\Psi_{B2}
\rangle + \gamma |\Psi_{B3} \rangle + \delta |\Psi_{B4} \rangle .
\] Then, the time evolution in the Bell basis is simply given by
\begin{eqnarray}
|\Psi (t) \rangle &=& \alpha |\Psi_{B1} \rangle + \beta |\Psi_{B2}
\rangle \\
&+& \gamma \Bigl[ \cos(\Delta t) |\Psi_{B3} \rangle - i
\sin(\Delta t) |\Psi_{B4} \rangle \Bigr] \nonumber \\ &+& \delta
\Bigl[ \cos(\Delta t) |\Psi_{B4} \rangle - i \sin(\Delta t)
|\Psi_{B3} \rangle \Bigr] . \nonumber
\end{eqnarray}} \item{Apply Hamiltonian evolution with a randomly
chosen time.} \item{Transform back to the measurement basis.}
\item{Do a parity measurement:
\begin{itemize}
\item{Find the probability of getting the result even ($E$), or
odd ($O$) from the state.} \item{Use these probabilities to choose
a random outcome, $E$ or $O$.} \item{Based on the result, update
the state.}
\end{itemize}}
\item{Transform back to the Bell basis.} \item{Repeat the
algorithm from step 2.}
\end{enumerate}
As mentioned before, this algorithm gives three possible outcomes:
(i) the parity meter measures $O$ all the time $\rightarrow$ state
$|\Psi_{\rm B1} \rangle$ has been prepared, (ii) the parity meter
measures $E$ all the time $\rightarrow$ state $|\Psi_{\rm B2}
\rangle$ has been prepared, and (iii) the parity meter measures a
string of a mixture of $O$ and $E$ results $\rightarrow$ a
dynamical superposition of the states $|\Psi_{\rm B3} \rangle$ and
$|\Psi_{\rm B4} \rangle$ has been prepared, which is not a steady
state of $H_{\rm SP}$. A long sequence of either $O$'s or $E$'s,
corresponding to cases (i) or (ii), indicates a statistically
confident preparation of a Bell state.

A statistical analysis of our model shows that if we start with a
product state in one of the two parity classes, e.g. $| \uparrow
\downarrow \rangle$ in the odd class, which can be easily prepared
experimentally, then the parity meter drives the system with
probability $1/2$ into the Bell state $|\Psi_{\rm B1} \rangle$.
(The same holds for the other parity class and the Bell state
$|\Psi_{\rm B2} \rangle$.)

If we, however, start with a random state, e.g. a fully mixed
state, then the parity meter still accomplishes a Bell state
preparation of the two states $|\Psi_{\rm B1} \rangle$ and
$|\Psi_{\rm B2} \rangle$ with a success probability of $1/4$ each.
If there are non-ideal symmetries, e.g. $\Delta_1 \neq \Delta_2$,
then, on longer time scales, there will be random switching
between the different parity classes.

Before proceeding to the next section, we briefly note that the
needed symmetry in the coupling constants between detector and
each of the qubits may be tested by DC current measurements, using
gate voltages to force the quantum dots into each of the four
classical configurations.  If there is any asymmetry, this will
show up in a slight current difference when comparing the
different configurations. The difference of the couplings
constants can by slightly tuned with the use of top gates.

\section{Applications of the parity meter}
\label{sec_bell}

We describe two applications of the parity meter of interest for
quantum information processing. The first one is a new proposal to
test Bell's inequality in the solid state. The second one is an
example of a realization of a CNOT gate using QPC's and DQD's as
building blocks.

\subsection{Testing Bell's inequality}

A slight modification of our setup as schematically shown in
Fig.~\ref{bmeas} can be used to violate Bell's inequality. The
Bell's inequality measurement consists essentially of four
consecutive steps:

\begin{figure}
\vspace{0.3cm}
\begin{center}
\epsfig{file=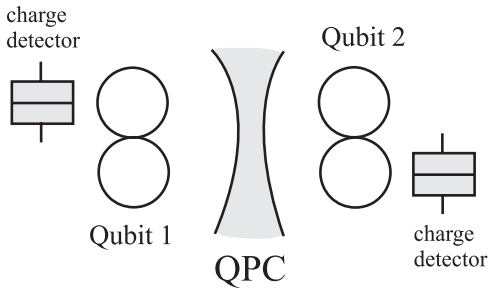,scale=0.7,=}
\caption{\label{bmeas} {\it Bell inequality setup.} In order to be
able to do a measurement of a violation of a Bell inequality, we
use a QPC as a parity meter between two qubits to create a Bell
state. The two outer detectors are then used to projectively
measure charge in the $\sigma_z$ basis of each qubit.}
\end{center}
\end{figure}

\begin{enumerate}
\item{{\it {Preparation step.}} During the preparation step, the
two DQD's have to be held at the symmetry point, i.e. $\epsilon_1
= \epsilon_2 = 0$ and $\Delta_1 = \Delta_2 = \Delta$, where the
Hamiltonian $H_{\rm SP}$ describes the two-qubit system. A
measurement of the parity meter, i.e. the center QPC, is done. If
the measurement is either $I_{\rm odd}$ or $I_{\rm even}$, then we
know that either state $|\Psi_{B1}\rangle$ or state
$|\Psi_{B2}\rangle$, respectively, has successfully been
prepared.} \item{{\it {Decoupling step.}} Once we know that the
system is in one of the two Bell states $|\Psi_{B1}\rangle$ or
$|\Psi_{B2}\rangle$, we would like to turn the coupling between
the two qubits off. In our setup, the coupling between the two
DQD's can be minimized by setting $V_{\rm bias} = 0$ and by
pinching off the QPC.} \item{{\it {Single qubit rotation step.}}
In order to do a measurement of a violation of a Bell inequality
in the Clauser-Horne-Shimony-Holt (CHSH) form, \cite{chsh} we have
to rotate each qubit and afterwards measure qubit-qubit
correlators in different bases. The single qubit rotation can be
done by pulsing $\Delta_1(t)$ and $\Delta_2(t)$ independently of
each other. Thus, during this step, we drive the system away from
the symmetry point. Note that this is the only option to measure
the two qubits in different bases, because typical measurement
devices for charge qubits can only measure in the $\sigma_z$ basis
of the qubit. Therefore, we have to rotate the state instead of
the measurement device (which is the usual practice in Bell
inequality measurements with photons).} \item{{\it {Measurement
step.}} Immediately after the {\it single qubit rotation step}, we
should be able to do a strong (projective) measurement in the
$\sigma_z$ basis of the qubit using high-fidelity single shot
detectors. This can, for instance, be accomplished by single
electron transistors as illustrated in Fig.~\ref{bmeas}. The time
delay between the two projective measurements is analogous to the
relative phase between the beam-splitters in the original CHSH
proposal.

An alternative to fast time resolved projective measurements
required for both the Bell inequality, and the CNOT gate of the
next section, is making a series of many weak ``kicked'' quantum
nondemolition measurements at a repetition rate commensurate with
the Rabi period of the qubits. In this alternative, the qubits are
not detuned from their symmetry point, and single-qubit rotations
are accomplished by simply waiting. \cite{jorda05}}
\end{enumerate}

We have to repeat the four steps many times with the same time
delay to obtain the correlation function
\begin{equation}
C_{\mathbf{a} \mathbf{b}} = \langle (\mathbf{a} \mathbf{\sigma})_1
\otimes (\mathbf{b} \mathbf{\sigma})_2 \rangle
\end{equation}
of the direct product of a ``spin'' measurement in qubit $1$ along
unit vector $\mathbf{a}$ and in qubit $2$ along unit vector
$\mathbf{b}$. Note that in our proposal the different angles
$\mathbf{a}$, $\mathbf{a'}$, $\mathbf{b}$, and $\mathbf{b'}$ are
realized by an appropriate application of the {\it single qubit
rotation step}. According to Bell, correlations are nonclassical
if we violate the inequality
\begin{equation}
{\cal B} = |C_{\mathbf{a} \mathbf{b}} + C_{\mathbf{a'} \mathbf{b}}
+ C_{\mathbf{a} \mathbf{b'}} - C_{\mathbf{a'} \mathbf{b'}}| \leq 2
.
\end{equation}
A simple way of analyzing the dephasing time in the system would
be to choose different time delays after which the {\it {single
qubit rotation step}} sets in.

\subsection{CNOT gate}
\label{sec_cnot}

\begin{figure}
\vspace{0.3cm}
\begin{center}
\epsfig{file=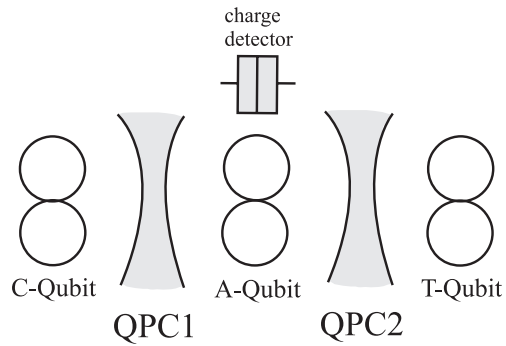,scale=0.65,=} \caption{\label{cnot}
{\it CNOT gate setup.} This setup contains three qubits and two
parity meters, i.e. QPC's, in between them. The qubits are the
control qubit (C-Qubit), the ancilla qubit (A-Qubit) and the
target qubit (T-Qubit). An additional projective measurement
device is attached to the A-qubit.}
\end{center}
\end{figure}

The setup of a realization of a CNOT gate using charge parity
meters is shown in Fig.~\ref{cnot} and follows the idea of
Ref.~\onlinecite{beena04}. It consists of three different charge
qubits and two parity meters. During the operation time of the
CNOT gate $t_{\rm op}$ coherent evolution of the charge qubits
should be negligible, which means that $\Delta t_{\rm op} \ll 1$.
Furthermore, $t_{\rm op}$ has to be smaller than the typical
dephasing time $T_2$ of the qubits. Note that in superconducting
charge qubits, a CNOT gate operation has already been demonstrated
experimentally \cite{yamam03} and interesting proposals for the
implementation of different kinds of two-qubit gates exist. \cite{wei05}

The Hamiltonian of the CNOT gate may be written as
\begin{equation}
H_{\rm CNOT} = H_{\rm QD} + H_{\rm QPC} + H_{\rm int}
\end{equation}
with
\begin{eqnarray}
H_{\rm QD} &=& -\frac{1}{2} \sum_{\alpha = 1,2,3} \Bigl(
\epsilon_\alpha \sigma_z^{(\alpha)} + \Delta_\alpha
\sigma_x^{(\alpha)} \Bigr) , \\
H_{\rm QPC} &=& H_{\rm QPC,1} + H_{\rm QPC,2} , \\
H_{\rm int} &=& \frac{\Delta \hat{E}_{1}}{2} \sigma_z^{(1)}
\sigma_z^{(2)} + \frac{\Delta \hat{E}_{2}}{2} \sigma_z^{(2)}
\sigma_z^{(3)} .
\end{eqnarray}
The energies $\epsilon_\alpha$ and $\Delta_\alpha$ ($\alpha =
1,2,3$) can (in principle) be controlled at any time by changing
the gate voltages that determine the single particle levels and
the tunnel couplings in each DQD system, respectively.

We now assume that the control qubit and the target qubit are in a
given state. It is important for the scheme to work that the
ancilla qubit is prepared in the state $|\Psi_{\rm AQ} \rangle
=(|\uparrow \rangle + |\downarrow\rangle)/\sqrt{2}$. It should be
rather easy to prepare the desired state of the ancilla qubit
because it is the ground state of the DQD system when the tunnel
coupling is finite. \cite{wiel03} Under the constraint $\Delta
t_{\rm op} \ll 1$, mentioned above, the outcome of the parity
measurement can just take two values, corresponding to $I_{\rm
odd}$ or $I_{\rm even}$ of Sec.~\ref{pent_sec}. Then, a
measurement of $I_{\rm odd}$ shows that the two qubits involved in
that measurement exist in the subspace spanned by the states
$|\uparrow \downarrow \rangle$ and $|\downarrow \uparrow \rangle$,
whereas a measurement of $I_{\rm even}$ indicates that the two
qubits involved in that measurement exist in the subspace spanned
by the states $|\uparrow \uparrow \rangle$ and $|\downarrow
\downarrow \rangle$.

Apart from the measurements of the two parity meters, the only
other ingredients needed to build a deterministic CNOT gate are
single qubit Hadamard gates ${\cal H} =(\sigma_x +
\sigma_z)/\sqrt{2}$. A Hadamard gate acting on qubit $\alpha$ can
be realized by tuning the corresponding single qubit Hamiltonian
\begin{equation}
{\cal H}_{\alpha} = -\frac{1}{2} \Bigl( \epsilon_\alpha
\sigma_z^{(\alpha)} + \Delta_\alpha \sigma_x^{(\alpha)} \Bigr)
\end{equation}
to the special symmetry point $\epsilon_\alpha = \Delta_\alpha$
and letting it act for a time $t_H = \pi/\sqrt{2} \Delta$. For
this single qubit rotation, we need to raise $\Delta$ and
$\epsilon$ temporarily.

The operation scheme of the CNOT gate goes as follows:
\begin{enumerate}
\item{{\it Preparation of the ancilla qubit.} This can be either
done by acting with ${\cal H}_2$ on state $|0\rangle$ of qubit 2
or by ground state preparation of a tunnel-coupled double dot.}
\item{{\it Parity measurement with QPC $1$.} The outcome of that
measurement $p_1$ has to be stored. $p_1=0$ corresponds to the odd
parity class, whereas $p_1=1$ corresponds to the even parity
class. The same holds, of course, for $p_2$, i.e. the outcome of
the parity measurement $2$ below. Afterwards, we need to decouple
the control and the ancilla qubit.} \item{{\it Hadamard step 1.}
Acting with ${\cal H}_2$ on the ancilla qubit and with ${\cal
H}_3$ on the target qubit, which means in practice to wait for an
appropriate time $t_H$ after tuning the single qubit
Hamiltonians.} \item{{\it Parity measurement with QPC $2$.} The
outcome of that measurement $p_2$ has to be stored. Afterwards, we
need to decouple the ancilla and the target qubit.} \item{{\it
Hadamard step 2.} Once more, acting with ${\cal H}_2$ on the
ancilla qubit and with ${\cal H}_3$ on the target qubit.}
\item{{\it Measurement of ancilla qubit.} This measurement has to
be done in a projective way.} \item{{\it Post-processing step.}
Depending on the outcome of the measured state of the ancilla
qubit as well as $p_1$ and $p_2$, we finally have to apply single
qubit operations to the control and the target qubit, which we
call $\sigma_c$ and $\sigma_t$. For the control qubit, $\sigma_c =
\sigma_z$ if $p_2 = 0$, while no post-processing of the control
qubit is needed if $p_2 = 1$. For the target qubit, $\sigma_t =
\sigma_x$ if the ancilla qubit is down and $p_1 = 1$, or if the
ancilla qubit is up and $p_1=0$. Otherwise, no post-processing of
the target qubit is needed. Applying a conditional operation of
$\sigma_x$ or $\sigma_z$, this means in practice making
$\Delta_\alpha \gg \epsilon_\alpha$ for the former case or vice
versa for the latter one.}
\end{enumerate}
As demonstrated in detail in Ref.~\onlinecite{beena04_cm}, the
different steps described above allow for a CNOT operation on the
control and the target qubit.

\section{Conclusions}
\label{sec_con}

We have proposed a realization of a charge parity meter, i.e. a
device that can distinguish between the subspaces of two parity
classes of quantum states but cannot distinguish between different
states in each parity class. If the states are two-qubit states
(in our case, the states that characterize two charge qubits) a
parity meter based on a QPC placed in a proper way between the two
qubits acts as an entangler just by a current measurement. Such a
device is a specific realization of a mesoscopic quadratic quantum
measurement. \cite{mao04}

Furthermore, we have demonstrated that the charge parity meter
supplemented by two single qubit charge detectors can be used to
do a measurement of a violation of Bell's inequality in the solid
state. To accomplish this, we have exploited the idea to use
single qubit rotations instead of a rotation of the measurement
device in order to be able to measure CHSH correlators in four
different bases.

Finally, a CNOT
gate operation has been described using two parity meters and
three qubits. Thereby, one of the three qubits just acts as an
ancilla qubit to enable a deterministic gate, whereas the other
two qubits are the standard control and target qubit that are
manipulated by the CNOT operation.

\begin{acknowledgments}

Interesting discussions with N. Gisin, F.H.L. Koppens, L.P.
Kouwenhoven, and L.M.K. Vandersypen are gratefully acknowledged.
A.N.J. thanks the Kavli Institute of Nanoscience at the TU Delft
for the kind hospitality. Financial support was provided by the
Dutch Science Foundation NWO/FOM, the Swiss NSF, and MaNEP. We
acknowledge support by the EC's Marie Curie Research Training
Network under contract MRTN-CT-2003-504574, Fundamentals of
Nanoelectronics. The work of
A.N.J. has been supported by the following grants AFRL
Grant No. F30602-01-1-0594, AFOSR Grant No. FA9550-04-1-0206, and
TITF Grant No. 2001-055.

\end{acknowledgments}


\begin{thebibliography}{10}

\bibitem{beena04}
C.W.J. Beenakker, D.P. DiVincenzo, C. Emary, and M. Kindermann,
Phys. Rev. Lett. {\bf 93}, 020501 (2004).

\bibitem{engel05}
H.-A. Engel and D. Loss, Science {\bf 309}, 586 (2005).

\bibitem{egues05}
For a perspective on these developments, see J. Carlos Egues,
Science {\bf 309}, 565 (2005).

\bibitem{mao04}
W. Mao, D.V. Averin, R. Ruskov, and A.N. Korotkov, Phys. Rev.
Lett. {\bf 93}, 056803 (2004).

\bibitem{rusko03}
R. Ruskov and A.N. Korotkov, Phys. Rev. B {\bf 67}, 241305(R)
(2003).

\bibitem{hayas03}
T. Hayashi, T. Fujisawa, H.D. Cheong, Y.H. Jeong, and Y. Hirayama,
Phys. Rev. Lett. {\bf 91}, 226804 (2003).

\bibitem{dqd_theo}
S. Gurvitz, Phys. Rev. B {\bf 56}, 15215 (1997); A.N. Korotkov,
{\it ibid.} {\bf 63}, 085312 (2001); A.N. Korotkov and D.V.
Averin, {\it ibid.} {\bf 64}, 165310 (2001); S. Pilgram and M.
B{\"u}ttiker, Phys. Rev. Lett. {\bf 89}, 200401 (2002); A.A.
Clerk, S.M. Girvin, and A.D. Stone, Phys. Rev. B {\bf 67}, 165324
(2003);  D.V. Averin and E.V. Sukhorukov, Phys. Rev. Lett. {\bf
95}, 126803 (2005); A.N. Jordan and M. B{\"u}ttiker, {\it ibid.}
{\bf 95}, 220401 (2005); A.A. Clerk, {\it ibid.} {\bf 96}, 056801 (2006).


\bibitem{dqd_exp}
J.R. Petta, A.C. Johnson, C.M. Marcus, M.P. Hanson, and A.C.
Gossard, Phys. Rev. Lett. {\bf 93}, 186802 (2004); J.M. Elzerman,
R. Hanson, L.H. Willems van Beveren, B. Witkamp, L.M.K.
Vandersypen, and L.P. Kouwenhoven, Nature {\bf 430}, 431 (2004);
J.R. Petta, A.C. Johnson, J.M. Taylor, E.A. Laird, A. Yacoby, M.D.
Lukin, C.M. Marcus, M.P. Hanson, and A.C. Gossard, Science {\bf
309}, 2180 (2005); F.H.L. Koppens, J.A. Folk, J.M. Elzerman, R.
Hanson, L.H. Willems van Beveren, I.T. Vink, H.P. Tranitz, W.
Wegscheider, L.P. Kouwenhoven, and L.M.K. Vandersypen, {\it ibid.}
{\bf 309}, 1346 (2005).

\bibitem{lambe06}
N. Lambert, R. Aguado, and T. Brandes, cond-mat/0602063.

\bibitem{voroj05}
S. Vorojtsov, E.R. Mucciolo, and H.U. Baranger, Phys. Rev. B {\bf
71}, 205322 (2005); Z.-J. Wu, K.-D. Zhu, X.-Z. Yuan, Y.-W. Jiang,
and H. Zheng, {\it ibid.} {\bf 71}, 205323 (2005).

\bibitem{itaku03}
T. Itakura and Y. Tokura, Phys. Rev. B {\bf 67}, 195320 (2003).

\bibitem{mason04}
N. Mason, M.J. Biercuk, and C.M. Marcus, Science {\bf 303}, 655 (2004).

\bibitem{gorma05}
J. Gorman, D.G. Hasko, and D.A. Williams, Phys. Rev. Lett. {\bf
95}, 090502 (2005).

\bibitem{rusko06}
R. Ruskov, A.N. Korotkov, and A. Mizel, Phys. Rev. B {\bf 73},
085317 (2006).

\bibitem{butti90}
M. B{\"u}ttiker, Phys. Rev. B {\bf 41}, R7906 (1990).

\bibitem{chris96}
T. Christen and M. B{\"u}ttiker, Phys. Rev. Lett. {\bf 77}, 143
(1996).

\bibitem{vande04}
L.M.K. Vandersypen, J.M. Elzerman, R.N. Schouten, L.H.Willems van
Beveren, R. Hanson, and L.P. Kouwenhoven, Appl. Phys. Lett. {\bf
85}, 4394 (2004).

\bibitem{chsh}
J.F. Clauser, M.A. Horne, A. Shimony, and R.A. Holt, Phys. Rev.
Lett. {\bf 23}, 880 (1969).

\bibitem{jorda05}
A.N. Jordan and M. B{\"u}ttiker, Phys. Rev. B {\bf 71}, 125333
(2005);  A.N. Jordan, A.N. Korotkov, and M. B{\"u}ttiker,
cond-mat/0510782.

\bibitem{yamam03}
T. Yamamoto, Yu.A. Pashkin, O. Astafiev, Y. Nakamura, and J.S.
Tsai, Nature {\bf 425}, 941 (2003).

\bibitem{wei05}
L.F. Wei, Yu-xi Liu, M.J. Storcz, and F. Nori, quant-ph/0508027.

\bibitem{wiel03}
W.G. van der Wiel, S. De Franceschi, J.M. Elzerman, T. Fujisawa,
S. Tarucha, and L.P. Kouwenhoven, Rev. Mod. Phys. {\bf 75}, 1
(2003).

\bibitem{beena04_cm}
C.W.J. Beenakker, D.P. DiVincenzo, C. Emary, and M. Kindermann,
quant-ph/0401066, App.~A.


\end{thebibliography}
\end{document}